\documentclass[dvips]{article}
\usepackage{supertabular,lscape,epsfig}
\usepackage{amssymb}
\usepackage{amsmath}

\DeclareSymbolFont{ppa}{OT1}{ppl}{m}{it}
\DeclareMathSymbol{\vv}{\mathalpha}{ppa}{'166}

\begin{document}

\newcommand{\dd}{\,{\rm d}}
\newcommand{\ie}{{\it i.e.},\,}
\newcommand{\etal}{{\it et al.\ }}
\newcommand{\eg}{{\it e.g.},\,}
\newcommand{\cf}{{\it cf.\ }}
\newcommand{\vs}{{\it vs.\ }}
\newcommand{\zdot}{\makebox[0pt][l]{.}}
\newcommand{\up}[1]{\ifmmode^{\rm #1}\else$^{\rm #1}$\fi}
\newcommand{\dn}[1]{\ifmmode_{\rm #1}\else$_{\rm #1}$\fi}
\newcommand{\upd}{\up{d}}
\newcommand{\uph}{\up{h}}
\newcommand{\upm}{\up{m}}
\newcommand{\ups}{\up{s}}
\newcommand{\arcd}{\ifmmode^{\circ}\else$^{\circ}$\fi}
\newcommand{\arcm}{\ifmmode{'}\else$'$\fi}
\newcommand{\arcs}{\ifmmode{''}\else$''$\fi}
\newcommand{\MS}{{\rm M}\ifmmode_{\odot}\else$_{\odot}$\fi}
\newcommand{\RS}{{\rm R}\ifmmode_{\odot}\else$_{\odot}$\fi}
\newcommand{\LS}{{\rm L}\ifmmode_{\odot}\else$_{\odot}$\fi}

\newcommand{\Abstract}[2]{{\footnotesize\begin{center}ABSTRACT\end{center}
\vspace{1mm}\par#1\par
\noindent
{~}{\it #2}}}

\newcommand{\TabCap}[2]{\begin{center}\parbox[t]{#1}{\begin{center}
  \small {\spaceskip 2pt plus 1pt minus 1pt T a b l e}
  \refstepcounter{table}\thetable \\[2mm]
  \footnotesize #2 \end{center}}\end{center}}

\newcommand{\TableSep}[2]{\begin{table}[p]\vspace{#1}
\TabCap{#2}\end{table}}

\newcommand{\FigCap}[1]{\footnotesize\par\noindent Fig.\  %
  \refstepcounter{figure}\thefigure. #1\par}

\newcommand{\TableFont}{\footnotesize}
\newcommand{\TableFontIt}{\ttit}
\newcommand{\SetTableFont}[1]{\renewcommand{\TableFont}{#1}}

\newcommand{\MakeTable}[4]{\begin{table}[htb]\TabCap{#2}{#3}
  \begin{center} \TableFont \begin{tabular}{#1} #4 
  \end{tabular}\end{center}\end{table}}

\newcommand{\MakeTableSep}[4]{\begin{table}[p]\TabCap{#2}{#3}
  \begin{center} \TableFont \begin{tabular}{#1} #4 
  \end{tabular}\end{center}\end{table}}

\newenvironment{references}%
{
\footnotesize \frenchspacing
\renewcommand{\thesection}{}
\renewcommand{\in}{{\rm in }}
\renewcommand{\AA}{Astron.\ Astrophys.}
\newcommand{\AAS}{Astron.~Astrophys.~Suppl.~Ser.}
\newcommand{\ApJ}{Astrophys.\ J.}
\newcommand{\ApJS}{Astrophys.\ J.~Suppl.~Ser.}
\newcommand{\ApJL}{Astrophys.\ J.~Letters}
\newcommand{\AJ}{Astron.\ J.}
\newcommand{\IBVS}{IBVS}
\newcommand{\PASP}{P.A.S.P.}
\newcommand{\Acta}{Acta Astron.}
\newcommand{\MNRAS}{MNRAS}
\renewcommand{\and}{{\rm and }}
\section{{\rm REFERENCES}}
\sloppy \hyphenpenalty10000
\begin{list}{}{\leftmargin1cm\listparindent-1cm
\itemindent\listparindent\parsep0pt\itemsep0pt}}%
{\end{list}\vspace{2mm}}

\def\TYLDA{~}
\newlength{\DW}
\settowidth{\DW}{0}
\newcommand{\dw}{\hspace{\DW}}

\newcommand{\refitem}[5]{\item[]{#1} #2%
\def\REFARG{#3}\ifx\REFARG\TYLDA\else, {\it#3}\fi
\def\REFARG{#4}\ifx\REFARG\TYLDA\else, {\bf#4}\fi
\def\REFARG{#5}\ifx\REFARG\TYLDA\else, {#5}\fi.}

\newcommand{\Section}[1]{\section{#1}}
\newcommand{\Subsection}[1]{\subsection{#1}}
\newcommand{\Acknow}[1]{\par\vspace{5mm}{\bf Acknowledgements.} #1}
\pagestyle{myheadings}

\newfont{\bb}{ptmbi8t at 12pt}
\newcommand{\xrule}{\rule{0pt}{2.5ex}}
\newcommand{\xxrule}{\rule[-1.8ex]{0pt}{4.5ex}}
\def\thefootnote{\fnsymbol{footnote}}
\begin{center}
{\Large\bf The Optical Gravitational Lensing Experiment.\\
\vskip3pt
{BVI} Maps of Dense Stellar Regions.\\
\vskip6pt
III. The Galactic Bulge\footnote{Based on  observations obtained
with the 1.3~m Warsaw telescope at the Las Campanas  Observatory of the
Carnegie Institution of Washington.}}

\vskip0.6cm
{\bf A.~~U~d~a~l~s~k~i$^1$,~~M.~~S~z~y~m~a~{\'n}~s~k~i$^1$,~~
M.~~K~u~b~i~a~k$^1$,\\ 
G.~~P~i~e~t~r~z~y~\'n~s~k~i$^{1,2}$,~~ I.~~S~o~s~z~y~\'n~s~k~i$^1$,~~
P.~~W~o~\'z~n~i~a~k$^3$,\\ K.~\.Z~e~b~r~u~\'n$^1$,
~~O.~~S~z~e~w~c~z~y~k$^1$~ and ~\L.~~W~y~r~z~y~k~o~w~s~k~i$^1$}
\vskip3mm
{$^1$Warsaw University Observatory, Al.~Ujazdowskie~4, 00-478~Warszawa,
Poland\\
e-mail: (udalski,msz,mk,pietrzyn,soszynsk,zebrun,szewczyk,wyrzykow)@astrouw.edu.pl\\
$^2$ Universidad de Concepci{\'o}n, Departamento de Fisica,
Casilla 160--C, Concepci{\'o}n, Chile\\
$^3$ Los Alamos National Laboratory, MS-D436, Los Alamos, NM 87545 USA\\
e-mail: wozniak@lanl.gov}
\end{center}

\Abstract{
We present the {\it VI} photometric maps of the Galactic bulge. They 
contain {\it VI} photometry and astrometry of about 30 million stars from 49 
fields of 0.225 square degree each in the Galactic center region. The data 
were collected  during the second phase of the OGLE microlensing project. We 
discuss the  accuracy of  data and present color-magnitude diagrams of 
selected fields  observed by OGLE in the Galactic bulge. 

The {\it VI} maps of the Galactic bulge are accessible electronically for the  
astronomical community from the OGLE Internet archive.}

\Section{Introduction}
Natural by-products of large microlensing surveys are huge photometric 
databa\-ses containing photometry of millions of stars from extremely interesting 
regions of the sky like the Galactic bulge, Galactic disk or Magellanic 
Clouds. These data may be used for many projects unrelated directly to 
microlensing. Long time baselines and very good accuracy of photometric data 
collected with modern CCD detectors  make them a gold mine for stellar studies
in both variable and non-variable domains. The list of possible projects is
long and actually not limited to star research (Dobrzycki \etal 2002, Eyer 
2002). 

The data collected during the second phase of the Optical Gravitational 
Lensing Experiment (OGLE-II, Udalski, Kubiak and Szyma{\'n}ski 1997) are 
particularly attractive for many projects requiring precise photometry. Among 
the other datasets collected during the microlensing searches, only the OGLE 
photometry was obtained with the standard {\it BVI} filters what makes it 
very well suited for many astronomical projects. Also, very good astronomical 
site where the OGLE project has been conducted -- the Las Campanas Observatory 
in Chile, made it possible to achieve the best resolution what is crucial in 
the very dense stellar fields like the Galactic bulge or Magellanic Clouds. 

In the previous papers of the series the OGLE project released the 
photometric maps of the Small and Large Magellanic Clouds (Udalski \etal 1998, 
Udalski \etal 2000) which contained the mean {\it BVI} photometry and 
astrometry of more than 7 million and 2 million stars from the central parts 
of the LMC and SMC, respectively. These maps have been widely used by many 
astronomers worldwide for many projects (\eg Zaritsky \etal 2002, Edge and Coe 
2002, Cordier \etal 2002, Subramaniam and Anupama 2002, Dobrzycki \etal 2002, 
Eyer 2002 and many more). 

In this paper, which is a continuation of the series, we release ``The OGLE 
{\it VI} Maps of the Galactic Bulge''. The maps contain the mean {\it VI} 
photometry and  astrometry of about 30 million stars from the fields covering 
about 11~square degrees in different parts of the Galactic center region.  

Because of potentially great impact on many astrophysical fields, in 
particular for studying poorly observed stellar populations toward the 
Galactic bulge, the OGLE policy is to make the photometric data available to 
the wide astronomical community. Similarly to our previous maps, the maps of 
the Galactic bulge are also available  from the OGLE Internet archive. Details 
are provided in the last Section of this paper. 

\Section{Observations} 
Observations presented in this paper were collected during the second phase of 
the OGLE microlensing search with the 1.3-m Warsaw telescope at Las Campanas 
Observatory, Chile. The observatory is operated by the Carnegie Institution  
of Washington. The telescope was equipped with the ``first generation'' camera  
with a SITe ${2048\times2048}$ CCD detector working in drift-scan mode. The  
pixel size was 24~$\mu$m giving the 0.417~arcsec/pixel scale. Observations  
were performed in the ``medium'' reading mode of the CCD detector with the gain  
7.1~e$^-$/ADU and readout noise of about 6.3~e$^-$. Details of the 
instrumentation setup can be found in Udalski, Kubiak and Szyma{\'n}ski 
(1997). 
\renewcommand{\arraystretch}{1.2}
\renewcommand{\TableFont}{\scriptsize}
\MakeTableSep{lccrr}{12.5cm}{OGLE-II fields in the Galactic bulge}
{\hline
\noalign{\vskip3pt}
\multicolumn{1}{c}{Field} & RA (J2000)  & DEC (J2000) & \multicolumn{1}{c}{$l$}
& \multicolumn{1}{c}{$b$} \\
\hline
\noalign{\vskip3pt}
 BUL$\_$SC1  & 18\uph02\upm32\zdot\ups5 & $-29\arcd57\arcm41\arcs$ &   $ 1\zdot\arcd08$ & $-3\zdot\arcd62$ \\
 BUL$\_$SC2  & 18\uph04\upm28\zdot\ups6 & $-28\arcd52\arcm35\arcs$ &   $ 2\zdot\arcd23$ & $-3\zdot\arcd46$ \\
 BUL$\_$SC3  & 17\uph53\upm34\zdot\ups4 & $-29\arcd57\arcm56\arcs$ &   $ 0\zdot\arcd11$ & $-1\zdot\arcd93$ \\
 BUL$\_$SC4  & 17\uph54\upm35\zdot\ups7 & $-29\arcd43\arcm41\arcs$ &   $ 0\zdot\arcd43$ & $-2\zdot\arcd01$ \\
 BUL$\_$SC5  & 17\uph50\upm21\zdot\ups7 & $-29\arcd56\arcm49\arcs$ &   $-0\zdot\arcd23$ & $-1\zdot\arcd33$ \\
 BUL$\_$SC6  & 18\uph08\upm03\zdot\ups7 & $-32\arcd07\arcm48\arcs$ &   $-0\zdot\arcd25$ & $-5\zdot\arcd70$ \\
 BUL$\_$SC7  & 18\uph09\upm10\zdot\ups6 & $-32\arcd07\arcm40\arcs$ &   $-0\zdot\arcd14$ & $-5\zdot\arcd91$ \\
 BUL$\_$SC8  & 18\uph23\upm06\zdot\ups2 & $-21\arcd47\arcm53\arcs$ &   $10\zdot\arcd48$ & $-3\zdot\arcd78$ \\
 BUL$\_$SC9  & 18\uph24\upm02\zdot\ups5 & $-21\arcd47\arcm55\arcs$ &   $10\zdot\arcd59$ & $-3\zdot\arcd98$ \\
 BUL$\_$SC10 & 18\uph20\upm06\zdot\ups6 & $-22\arcd23\arcm03\arcs$ &   $ 9\zdot\arcd64$ & $-3\zdot\arcd44$ \\
 BUL$\_$SC11 & 18\uph21\upm06\zdot\ups5 & $-22\arcd23\arcm05\arcs$ &   $ 9\zdot\arcd74$ & $-3\zdot\arcd64$ \\
 BUL$\_$SC12 & 18\uph16\upm06\zdot\ups3 & $-23\arcd57\arcm54\arcs$ &   $ 7\zdot\arcd80$ & $-3\zdot\arcd37$ \\
 BUL$\_$SC13 & 18\uph17\upm02\zdot\ups6 & $-23\arcd57\arcm44\arcs$ &   $ 7\zdot\arcd91$ & $-3\zdot\arcd58$ \\
 BUL$\_$SC14 & 17\uph47\upm02\zdot\ups7 & $-23\arcd07\arcm30\arcs$ &   $ 5\zdot\arcd23$ & $ 2\zdot\arcd81$ \\
 BUL$\_$SC15 & 17\uph48\upm06\zdot\ups9 & $-23\arcd06\arcm09\arcs$ &   $ 5\zdot\arcd38$ & $ 2\zdot\arcd63$ \\
 BUL$\_$SC16 & 18\uph10\upm06\zdot\ups7 & $-26\arcd18\arcm05\arcs$ &   $ 5\zdot\arcd10$ & $-3\zdot\arcd29$ \\
 BUL$\_$SC17 & 18\uph11\upm03\zdot\ups6 & $-26\arcd12\arcm35\arcs$ &   $ 5\zdot\arcd28$ & $-3\zdot\arcd45$ \\
 BUL$\_$SC18 & 18\uph07\upm03\zdot\ups5 & $-27\arcd12\arcm48\arcs$ &   $ 3\zdot\arcd97$ & $-3\zdot\arcd14$ \\
 BUL$\_$SC19 & 18\uph08\upm02\zdot\ups4 & $-27\arcd12\arcm45\arcs$ &   $ 4\zdot\arcd08$ & $-3\zdot\arcd35$ \\
 BUL$\_$SC20 & 17\uph59\upm19\zdot\ups1 & $-28\arcd52\arcm55\arcs$ &   $ 1\zdot\arcd68$ & $-2\zdot\arcd47$ \\
 BUL$\_$SC21 & 18\uph00\upm22\zdot\ups3 & $-28\arcd51\arcm45\arcs$ &   $ 1\zdot\arcd80$ & $-2\zdot\arcd66$ \\
 BUL$\_$SC22 & 17\uph56\upm47\zdot\ups6 & $-30\arcd47\arcm46\arcs$ &   $-0\zdot\arcd26$ & $-2\zdot\arcd95$ \\
 BUL$\_$SC23 & 17\uph57\upm54\zdot\ups5 & $-31\arcd12\arcm36\arcs$ &   $-0\zdot\arcd50$ & $-3\zdot\arcd36$ \\
 BUL$\_$SC24 & 17\uph53\upm17\zdot\ups9 & $-32\arcd52\arcm45\arcs$ &   $-2\zdot\arcd44$ & $-3\zdot\arcd36$ \\
 BUL$\_$SC25 & 17\uph54\upm26\zdot\ups1 & $-32\arcd52\arcm49\arcs$ &   $-2\zdot\arcd32$ & $-3\zdot\arcd56$ \\
 BUL$\_$SC26 & 17\uph47\upm15\zdot\ups5 & $-34\arcd59\arcm31\arcs$ &   $-4\zdot\arcd90$ & $-3\zdot\arcd37$ \\
 BUL$\_$SC27 & 17\uph48\upm23\zdot\ups6 & $-35\arcd09\arcm32\arcs$ &   $-4\zdot\arcd92$ & $-3\zdot\arcd65$ \\
 BUL$\_$SC28 & 17\uph47\upm05\zdot\ups8 & $-37\arcd07\arcm47\arcs$ &   $-6\zdot\arcd76$ & $-4\zdot\arcd42$ \\
 BUL$\_$SC29 & 17\uph48\upm10\zdot\ups8 & $-37\arcd07\arcm21\arcs$ &   $-6\zdot\arcd64$ & $-4\zdot\arcd62$ \\
 BUL$\_$SC30 & 18\uph01\upm25\zdot\ups0 & $-28\arcd49\arcm55\arcs$ &   $ 1\zdot\arcd94$ & $-2\zdot\arcd84$ \\
 BUL$\_$SC31 & 18\uph02\upm22\zdot\ups6 & $-28\arcd37\arcm21\arcs$ &   $ 2\zdot\arcd23$ & $-2\zdot\arcd94$ \\
 BUL$\_$SC32 & 18\uph03\upm26\zdot\ups8 & $-28\arcd38\arcm02\arcs$ &   $ 2\zdot\arcd34$ & $-3\zdot\arcd14$ \\
 BUL$\_$SC33 & 18\uph05\upm30\zdot\ups9 & $-28\arcd52\arcm50\arcs$ &   $ 2\zdot\arcd35$ & $-3\zdot\arcd66$ \\
 BUL$\_$SC34 & 17\uph58\upm18\zdot\ups5 & $-29\arcd07\arcm50\arcs$ &   $ 1\zdot\arcd35$ & $-2\zdot\arcd40$ \\
 BUL$\_$SC35 & 18\uph04\upm28\zdot\ups6 & $-27\arcd56\arcm56\arcs$ &   $ 3\zdot\arcd05$ & $-3\zdot\arcd00$ \\
 BUL$\_$SC36 & 18\uph05\upm31\zdot\ups2 & $-27\arcd56\arcm44\arcs$ &   $ 3\zdot\arcd16$ & $-3\zdot\arcd20$ \\
 BUL$\_$SC37 & 17\uph52\upm32\zdot\ups2 & $-29\arcd57\arcm44\arcs$ &   $ 0\zdot\arcd00$ & $-1\zdot\arcd74$ \\
 BUL$\_$SC38 & 18\uph01\upm28\zdot\ups0 & $-29\arcd57\arcm01\arcs$ &   $ 0\zdot\arcd97$ & $-3\zdot\arcd42$ \\
 BUL$\_$SC39 & 17\uph55\upm39\zdot\ups1 & $-29\arcd44\arcm52\arcs$ &   $ 0\zdot\arcd53$ & $-2\zdot\arcd21$ \\
 BUL$\_$SC40 & 17\uph51\upm06\zdot\ups1 & $-33\arcd15\arcm11\arcs$ &   $-2\zdot\arcd99$ & $-3\zdot\arcd14$ \\
 BUL$\_$SC41 & 17\uph52\upm07\zdot\ups2 & $-33\arcd07\arcm41\arcs$ &   $-2\zdot\arcd78$ & $-3\zdot\arcd27$ \\
 BUL$\_$SC42 & 18\uph09\upm05\zdot\ups0 & $-26\arcd51\arcm53\arcs$ &   $ 4\zdot\arcd48$ & $-3\zdot\arcd38$ \\
 BUL$\_$SC43 & 17\uph35\upm13\zdot\ups5 & $-27\arcd11\arcm00\arcs$ &   $ 0\zdot\arcd37$ & $ 2\zdot\arcd95$ \\
 BUL$\_$SC44 & 17\uph49\upm22\zdot\ups4 & $-30\arcd02\arcm45\arcs$ &   $-0\zdot\arcd43$ & $-1\zdot\arcd19$ \\
 BUL$\_$SC45 & 18\uph03\upm36\zdot\ups5 & $-30\arcd05\arcm00\arcs$ &   $ 0\zdot\arcd98$ & $-3\zdot\arcd94$ \\    
 BUL$\_$SC46 & 18\uph04\upm39\zdot\ups7 & $-30\arcd05\arcm11\arcs$ &   $ 1\zdot\arcd09$ & $-4\zdot\arcd14$ \\
 BUL$\_$SC47 & 17\uph27\upm03\zdot\ups7 & $-39\arcd47\arcm16\arcs$ &  $-11\zdot\arcd19$ & $-2\zdot\arcd60$ \\
 BUL$\_$SC48 & 17\uph28\upm14\zdot\ups0 & $-39\arcd46\arcm58\arcs$ &  $-11\zdot\arcd07$ & $-2\zdot\arcd78$ \\
 BUL$\_$SC49 & 17\uph29\upm25\zdot\ups1 & $-40\arcd16\arcm21\arcs$ &  $-11\zdot\arcd36$ & $-3\zdot\arcd25$ \\
\hline}
\renewcommand{\TableFont}{\small}

Forty nine driftscan fields covering about 11 square degrees were monitored 
regularly (practically on every clear night) in the observing seasons 
1997--2000. Each of the fields covers $14.2\times57$~arcmin on the sky. Table~1 
lists the equatorial coordinates of the center of each field and field
acronym. Positions of some fields overlap by about one arcmin for testing 
purposes. The driftscan fields were selected to sample wide range (${-
12\arcd<l<12\arcd}$) of the Galactic bulge regions. They were located typically 
in the strip of lower interstellar extinction below the Galactic equator at 
${b\approx-3\arcd}$ in the areas of the highest stellar density. At positive 
galactic latitude low extinction windows are rare so only a few fields where 
the stellar density is high enough for microlensing search were selected. 
Fig.~1 presents the schematic map of the Galactic bulge with location of all 
OGLE-II bulge fields. 
\begin{figure}[htb]
\vglue-21mm
\centerline{\includegraphics[width=12.7cm]{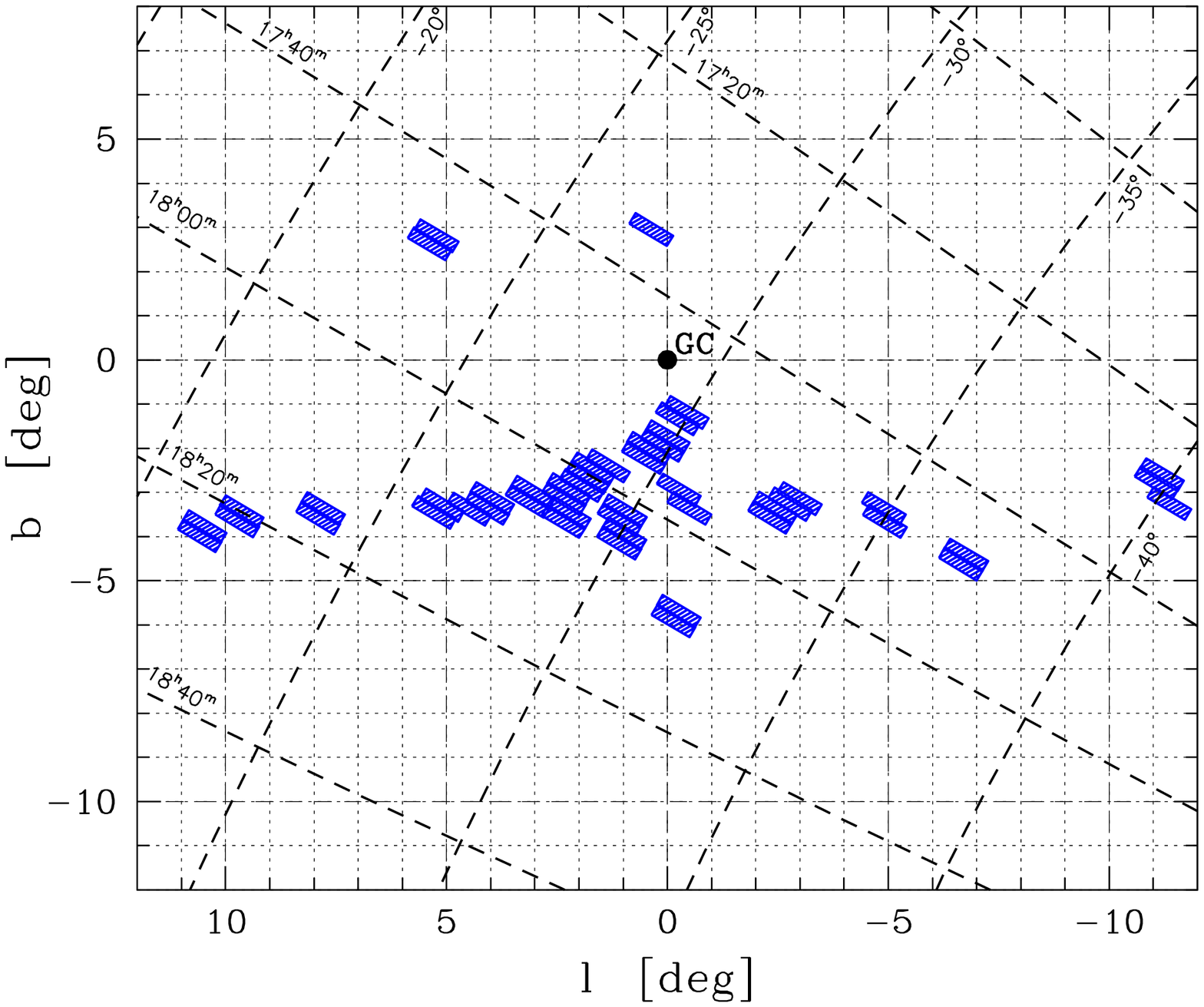}}
\FigCap{OGLE-II fields in the Galactic bulge.}
\end{figure}%1

Regular observations of the Galactic bulge fields started on March 23, 1997 
and continued up to November 23, 2000. Observations of three fields BUL$\_$SC47, 
BUL$\_$SC48 and BUL$\_$SC49 began one year later in 1998 observing season. Two 
fields, namely BUL$\_$SC45 and BUL$\_$SC46 were observed less frequently 
mainly to  maintain phasing of variable stars discovered in these fields 
during the OGLE-I phase. Photometry of part of BUL$\_$SC45 field, based on 
much smaller observational material has already been presented in 
Paczy{\'n}ski \etal (1999). 

Observations were obtained with the standard {\it VI} filters closely 
reproducing the standard system (Section~3). Due to the microlensing search  
observing strategy, the vast majority of observations were obtained through  
the {\it I}-band filter (150--570 per field) while much smaller number of  
frames in the {\it V}-band were collected (5--18). The effective exposure time 
was 87 and 124~seconds, for the {\it I} and {\it V}-band, respectively. 
After August 14, 1998 (HJD=2451040) the exposure time for the {\it I}-band was 
increased to 99.3 seconds. 

More than 16~000 images (about 540~GB of raw data) of the Galactic bulge 
fields were collected during the OGLE-II phase of the OGLE survey. Because of  
high stellar density of the fields, observations were only conducted during 
the nights with good seeing conditions. The median seeing of the entire data 
set is about 1.25~arcsec. Observations were usually suspended when the seeing 
exceeded 1.6--1.8~arcsec. 

\Section{Data Reduction and Calibration}
All collected frames were reduced using the standard OGLE data pipeline in the  
identical manner as the SMC and LMC data (Udalski \etal 1998, Udalski \etal 
2000). The data pipeline is  described in detail in Udalski \etal (1998). 

To summarize, after de-biasing and flat-fielding, photometry of objects was 
derived using the {\sc DoPhot}  photometry program (Schechter, Saha and Mateo 
1993) running in the fixed position  mode on sixty four ${512\times 
512}$~pixel subframes. The driftscan image (${2048\times8192}$~pixels) was 
first matched with the so  called ``template'' image, \ie the image with very 
good angular resolution (obtained at very good seeing conditions) and then 
divided into subframes. Photometry of each subframe was tied to the photometry 
of the template subframe  by adding the mean shift derived usually from 
several hundreds bright stars. Thus, photometry of the template image 
defines the instrumental system. Objects  detected in the template image for 
the {\it V}-band were first matched with the {\it I}-band template image 
objects of a given field so the star numbering was the same in all bands 
making the data handling much easier. Due to small shifts of the {\it V} 
template images in respect to the {\it I}-band image not all stars detected 
in the {\it I}-band image have  {\it V} photometry. 

To determine transformations of the instrumental photometry to the standard  
system, several Landolt (1992) standard fields were observed on about 250  
photometric nights during the OGLE-II observations. Based on thousands  
observations of standard stars in Landolt (1992) fields located all over the  
sky and covering large range of colors (${-0.2~{\rm mag}<V-I<2~{\rm mag}}$), 
the following average transformations were derived: 
\begin{eqnarray} 
V  &=&\vv-0.002\times(V-I) +{\rm const}_V\nonumber\\ 
I  &=&  i+0.029\times(V-I) +{\rm const}_I\\ 
V-I&=& 0.969\times(\vv-i)+{\rm const}_{V-I}\nonumber 
\end{eqnarray} 

The typical residuals of calculated minus observed magnitudes of standard  
stars did not exceed 0.03~mag. Observations of standard stars indicate that  
the instrumental system was extremely stable during the period of observations  
and the standard system magnitudes could be derived with good accuracy  
(${0.02-0.04}$~mag) even during the photometric nights when no standard stars 
were observed. 

Transformation of the instrumental magnitudes to the standard system was 
performed in  the following steps. First, the aperture corrections were 
determined on  each of the 64 subframes. They were derived from aperture 
photometry of  typically 20--100 stars per subframe measured in images with  
faint stars subtracted. Then the total correction consisting of the aperture 
correction, zero point of transformation, extinction correction and 
normalization to 1~sec exposure time was obtained. The total corrections 
were derived independently for about 5 and 30--40 photometric nights for 
the {\it V} and {\it I}-band, respectively. Typical standard deviation of the 
total correction in each of the 64 subframes was of about 0.02--0.03~mag. The 
mean values of the total correction were subsequently used for the 
construction of photometric databases (Szyma{\'n}ski and Udalski 1993, Udalski 
\etal 1998) for the {\it V} and {\it I}-band. The databases contain entire 
photometry of all objects in a given OGLE field in the system very close to 
the standard one -- only the color term (Eq.~1) is not included. 

Equatorial coordinates of objects detected in the OGLE fields were determined  
in the identical manner as described in Udalski \etal (1998). Objects in the  
OGLE-II frames were cross-identified with the objects detected in the  
Digitized Sky Survey images, and the transformation between the OGLE pixel  
grid and (RA,DEC) coordinates in the DSS coordinate system was derived.  
About 2000--14000 stars were used for transformation depending on the stellar  
density in the field. The internal accuracy of the determined equatorial  
coordinates, as measured in the overlapping regions of neighboring fields is  
about 0.15--0.20~arcsec. However we remind that the systematic error of the 
DSS coordinate  system may reach 0.7~arcsec. 

\Section{Accuracy of Transformation}
As we mentioned in the previous Section the OGLE instrumental photometric data 
were tied to the standard Landolt (1992) system. While the derived 
transformations (Eq.~1) indicate close approximation of the Landolt's system 
it should be noted that the color range of observed standard stars was limited 
to ${V-I<2}$~mag. For redder stars the transformations become extrapolation 
and are prone to systematic errors. While such systematic effects were 
negligible in the case of the OGLE-II maps of the Magellanic Clouds, because 
the vast majority of stars were bluer than ${V-I\approx2}$~mag the situation 
is different in the Galactic bulge case. Many Galactic bulge fields are 
severely reddened due to large interstellar extinction toward the Galactic 
center and in some of the fields even more than 50\% stars have ${V-I>2}$~mag. 

One of the main source of systematic errors could be different transmission of 
the OGLE filters compared to the standard system definition. While the {\it 
V}-band filter is a standard combo of Schott glass filters used to reproduce 
the {\it V}-band with CCD detectors (Bessell 1990) and it should not introduce 
any significant systematic errors, the {\it I}-band filter is our main 
concern. OGLE-II used 4-mm RG9 Schott glass filter for {\it I}-filter 
approximation which is an alternative to the usually used interference filters 
(Bessell 1990). The glass filter replacement has wider long-wavelength wing, 
that is defined by CCD detector sensitivity, contrary to the much sharper drop 
of transmission at about 9000~\AA~of the standard {\it I} pass-band defined 
either by interference filter or by fall of sensitivity of photomultipliers 
originally used for definition of the {\it I}-band (Bessell 1990, Landolt 
1992). Although this excess of red-side sensitivity of glass filter is rapidly 
falling with wavelength it can cause some non-linearities in transformations 
to the standard system, especially for very red objects or highly reddened by 
interstellar extinction regular stars. The latter are quite common in the 
Galactic bulge. 

Moreover, the excess of the red sensitivity of OGLE-II filter can be affected 
by atmospheric transmission. In the range of 9000--9900~\AA~the atmospheric 
transmission is reduced by telluric absorption bands. The amount of telluric 
absorption varies  with the zenith distance and the amount of precipitable 
water vapor. This effect makes the OGLE-II filter closer to the standard one, 
but on the other hand, more sensitive to atmospheric conditions. Fortunately, 
at very good observing site like Las Campanas Observatory the atmospheric 
conditions are very stable in long term scale and the large number of 
photometric nights when calibrations were performed averages atmospheric 
condition fluctuations. 

Ideally, one should determine differences between the OGLE-II and standard 
{\it I}-band filters by observing large number of red color standards. 
Unfortunately, the number of red standards is very limited and it is 
practically impossible to calibrate reliably the filters up to ${V-
I\approx4}$~mag. In practice, the reddest stars in the Landolt's (1992) 
standard fields observed by OGLE-II reached only ${V-I\approx2}$~mag. 
Therefore to estimate the possible systematic errors for very red stars we 
calculated and compared  expected {\it I}-band magnitudes for OGLE-II and 
standard filters using models of atmospheres by Kurucz (1993). 

\begin{figure}[h]
\vglue-7mm
\centerline{\includegraphics[width=11cm]{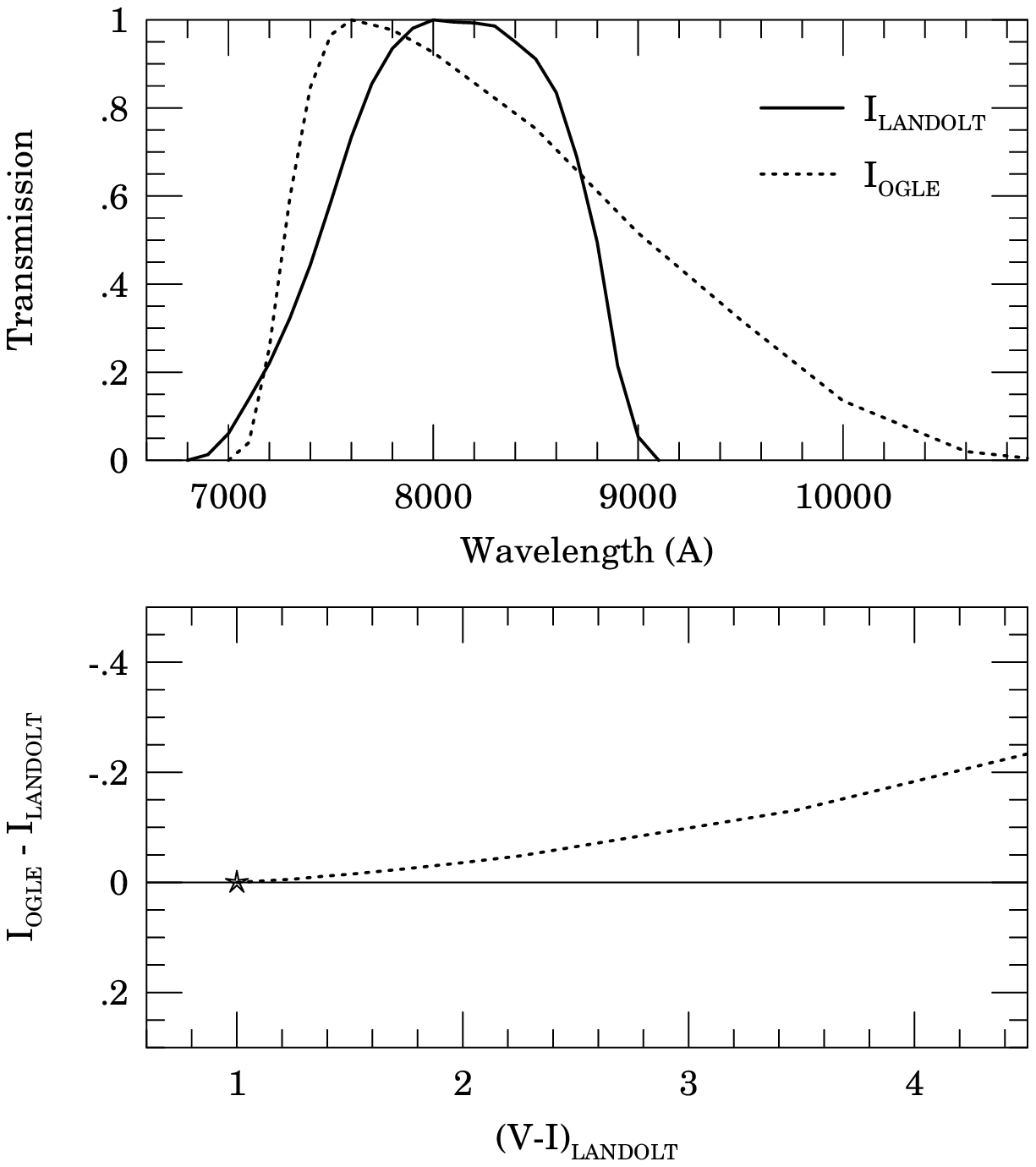}}
\FigCap{{\it Upper panel}: dotted line -- transmission of the OGLE-II 
{\it I}-band filter; solid line -- transmission of the standard {\it I}-band 
filter according to Landolt (1992). {\it Lower panel}: Difference between 
OGLE-II filter magnitudes transformed to the Landolt's system with Eq.~(1) and 
Landolt's standard {\it I}-band filter magnitudes (dotted line) for typical 
red clump giant (asterisk) reddened by standard extinction law.} 
\end{figure}%2
First, we derived the OGLE-II {\it I}-filter transmission curve using Schott 
glass filter catalog data and OGLE-II CCD detector quantum efficiency curve 
from manufacturer data sheet. It is shown in the upper panel of Fig.~2 with 
the transmission curve of the standard {\it I}-band filter as defined by 
Landolt (1992). We assumed full atmospheric transmission, so our further 
results are an upper limit of possible filter differences. As we mentioned 
above telluric absorption makes the filter transmissions more similar. It is 
known that both -- the filter transmission and CCD sensitivity -- can vary by 
a few percent from the catalog data. Therefore we additionally performed 
further calculations for transmission different by 5\% from that shown in 
Fig.~2. In both cases results were practically identical. 

We selected a typical red clump giant star for modeling. Such a star has ${V-
I\approx1.0}$~mag \ie its color is more or less in the middle of the range 
calibrated by standards (${V-I<2}$~mag). We took the Kurucz's model of 
atmosphere of a giant with metallicity of ${{\rm [Fe/H]}=-0.2}$ (typical for 
Galactic bulge), ${\log g=2.00}$ and ${T_{\rm eff}=4500}$~K -- typical red 
clump giant values. We used somewhat modified  {\sc cousins.for} program from 
the Kurucz's archive to determine the {\it I}-band magnitudes for such a giant 
star for OGLE-II and standard Landolt's {\it I}-band filter. Then we repeated 
calculations making the star redder by introducing the interstellar extinction 
obeying standard extinction law with ${R_V=3.1}$ (Cardelli, Clayton and Mathis 
1989). Calculations were performed up to very high reddening values of ${ E(B-
V)=3.0}$~mag. The OGLE-II filter results were then transformed to the 
Landolt's system using our empirical transformations given by Eq.~(1). 

Lower panel of Fig.~2 presents results of our calculations. Asterisk marks 
position of unreddened red clump giant. Dotted line indicates the calculated 
difference of the  OGLE-II filter values transformed to the Landolt's system 
with Eq.~(1) transformations and  magnitudes obtained for the Landolt's 
pass-band. As we expected, in the range covered by standard stars (${V-
I<2}$~mag) consistency of both magnitudes is very good at 0.03~mag level. 
However, for redder objects there is a systematic trend giving brighter {\it 
I}-band magnitudes (and redder ${V-I}$ colors) for the OGLE-II filter. This 
effect may reach 0.25~mag at very red ${V-I>4}$~mag stars and it is fully 
understandable, as for redder stars the red excess of the OGLE-II filter 
transmission is not fully compensated by linear transformation (Eq.~1) based 
on much bluer stars. 

In practice, the differences presented in the lower panel of Fig.~2 are likely 
to be an upper limit of possible discrepancy between the OGLE-II and Landolt's 
standard system. For instance, it is likely that due to telluric absorption 
the real discrepancy is smaller. Because we are not able to determine precise 
empirical transformation to the Landolt system in the entire range of observed 
colors in the Galactic bulge, we leave the data transformed with Eq.~(1). The 
reader should be, however, aware that for very red stars (${V-I}>2$~mag) the 
magnitudes presented in the OGLE-II maps of the Galactic bulge can differ from 
the Landolt's system values even as much as presented in Fig.~2. 

\Section{{\bb VI} Maps of the Galactic Bulge}
The {\it VI} maps of the Galactic bulge were constructed using the photometric 
databases  of each field. First, the mean magnitudes of each object were  
calculated with $5\sigma$ clipping alghoritm. Then, we corrected the  
magnitudes for a small systematic error, caused by non-perfect flat-fielding  
at the edges of the field. This effect was first noticed by Dr.\ D.S.\ Graff  
(Ohio State University), and it was precisely mapped based on observations of  
hundreds of standard stars. Finally, the color corrections (Eq.~1) were  
derived and added. Only objects with more than 50 good observations (see 
Udalski \etal 1998) in the {\it I}-band were included in the final maps of 
the Galactic bulge. The errors of zero points of photometry should not exceed 
0.04~mag. Table~2 lists the total number of objects in the  OGLE-II maps of 
the Galactic bulge fields. 
\MakeTable{lcclc}{12.5cm}{Number of objects in the OGLE-II Galactic bulge maps}
{\cline{1-2}\cline{4-5}
\noalign{\vskip3pt}
\multicolumn{1}{c}{Field} & $N_{\rm objects}$ &~~~~~~&
\multicolumn{1}{c}{Field} & $N_{\rm objects}$ \\
\noalign{\vskip3pt}
\cline{1-2}\cline{4-5}
\noalign{\vskip3pt}
BUL$\_$SC1  &  729852 &~~~~~~ &BUL$\_$SC26 &  728200\\  
BUL$\_$SC2  &  803269 &~~~~~~ &BUL$\_$SC27 &  690785\\  
BUL$\_$SC3  &  805587 &~~~~~~ &BUL$\_$SC28 &  405799\\  
BUL$\_$SC4  &  774091 &~~~~~~ &BUL$\_$SC29 &  491941\\  
BUL$\_$SC5  &  433990 &~~~~~~ &BUL$\_$SC30 &  762481\\  
BUL$\_$SC6  &  514084 &~~~~~~ &BUL$\_$SC31 &  790471\\  
BUL$\_$SC7  &  462748 &~~~~~~ &BUL$\_$SC32 &  797493\\  
BUL$\_$SC8  &  401813 &~~~~~~ &BUL$\_$SC33 &  738508\\  
BUL$\_$SC9  &  330338 &~~~~~~ &BUL$\_$SC34 &  960656\\  
BUL$\_$SC10 &  458816 &~~~~~~ &BUL$\_$SC35 &  770940\\  
BUL$\_$SC11 &  425984 &~~~~~~ &BUL$\_$SC36 &  873472\\  
BUL$\_$SC12 &  534720 &~~~~~~ &BUL$\_$SC37 &  664424\\  
BUL$\_$SC13 &  569850 &~~~~~~ &BUL$\_$SC38 &  710234\\  
BUL$\_$SC14 &  619028 &~~~~~~ &BUL$\_$SC39 &  784316\\  
BUL$\_$SC15 &  600787 &~~~~~~ &BUL$\_$SC40 &  630774\\  
BUL$\_$SC16 &  699804 &~~~~~~ &BUL$\_$SC41 &  603404\\  
BUL$\_$SC17 &  687019 &~~~~~~ &BUL$\_$SC42 &  600519\\  
BUL$\_$SC18 &  749265 &~~~~~~ &BUL$\_$SC43 &  474367\\  
BUL$\_$SC19 &  732089 &~~~~~~ &BUL$\_$SC44 &  318561\\  
BUL$\_$SC20 &  785317 &~~~~~~ &BUL$\_$SC45 &  627412\\  
BUL$\_$SC21 &  882518 &~~~~~~ &BUL$\_$SC46 &  551815\\  
BUL$\_$SC22 &  715301 &~~~~~~ &BUL$\_$SC47 &  300705\\  
BUL$\_$SC23 &  723687 &~~~~~~ &BUL$\_$SC48 &  286907\\  
BUL$\_$SC24 &  612189 &~~~~~~ &BUL$\_$SC49 &  251629\\  
BUL$\_$SC25 &  622326 &~~~~~~ &            &        \\  
\cline{1-2}\cline{4-5}
\noalign{\vskip3pt}     
&&& Total: & 30490285}                 

Table~3 presents the sample data from the map of the BUL$\_$SC1 field. In the  
consecutive columns the following data are provided: star ID number,  
equatorial coordinates, ($X,Y$) coordinates in the {\it I}-band template 
image,  photometry: {\it V}, ${(V-I)}$, {\it I}, number of observations, 
number of rejected observations and standard deviation for the {\it VI}-bands, 
respectively. In the electronic version we additionally  provide the FITS 
template images for easy object identification. Maps of all Galactic bulge 
fields are available electronically from the OGLE Internet archive (see 
Section~8). 

It is worth noting that although the maps contain only the mean photometry of 
detected stars they can also be used for discrimination of variable stars. 
Large standard deviation of {\it I}-band magnitudes usually indicates stellar 
variability. It should be also noted that the preliminary catalog of about 
200~000 variable stars found in the OGLE-II Galactic bulge fields has already 
been released (Wo{\'z}niak \etal 2002; it is also available from the OGLE 
Internet archive). The maps can be used for more accurate calibration of light 
curves from that catalog. As the numbering of objects is different in both 
catalogs, cross-identification can be done by comparison of pixel  or 
equatorial coordinates  (the pixel coordinates system is the same in both 
catalogs). 

\begin{landscape}
\renewcommand{\arraystretch}{.85}
\MakeTable{c@{\hspace{9pt}}c@{\hspace{9pt}}c@{\hspace{12pt}}r
@{\hspace{12pt}}r@{\hspace{12pt}}c@{\hspace{12pt}}c@{\hspace{12pt}}c
@{\hspace{12pt}}r@{\hspace{9pt}}c@{\hspace{9pt}}c
@{\hspace{9pt}}r@{\hspace{9pt}}c@{\hspace{9pt}}c}{12.5cm}{Sample 
of data from the {\it VI} map of the field BUL$\_$SC1}
{
\hline
\noalign{\vskip5pt}
Star & RA & DEC & \multicolumn{1}{c}{$X$} & \multicolumn{1}{c}{$Y$} & $V$ & $V-I$ & 
$I$ & $N_{\rm ok}^V$ & $N_{\rm bad}^V$ & $\sigma_V$ & $N_{\rm ok}^I$ & $N_{\rm bad}^I$ & $\sigma_I$\\
no & (J2000) & (J2000) &&&&&&&&&&&\\
\noalign{\vskip5pt}
\hline
\noalign{\vskip5pt}

 1 & 18\uph02\upm13\zdot\ups55 & $-30\arcd25\arcm52\zdot\arcs6$ &  436.71 &   6.63 & 14.702 &  1.842 & 12.857 &   9 &  0 & 0.011 & 134 &  1 & 0.018 \\
 2 & 18\uph02\upm13\zdot\ups25 & $-30\arcd25\arcm50\zdot\arcs3$ &  427.20 &  12.09 & 14.272 &  1.845 & 12.423 &  11 &  0 & 0.011 & 153 &  2 & 0.014 \\
 3 & 18\uph02\upm07\zdot\ups46 & $-30\arcd25\arcm25\zdot\arcs4$ &  245.93 &  72.10 & 14.430 &  2.350 & 12.076 &  10 &  0 & 0.012 & 218 &  0 & 0.014 \\
 4 & 18\uph02\upm05\zdot\ups04 & $-30\arcd25\arcm24\zdot\arcs8$ &  170.42 &  73.28 & 14.597 &  2.206 & 12.388 &  10 &  0 & 0.014 & 205 &  0 & 0.013 \\
 5 & 18\uph02\upm06\zdot\ups55 & $-30\arcd25\arcm02\zdot\arcs5$ &  217.51 & 127.31 & 16.786 &  4.338 & 12.440 &  11 &  0 & 0.123 & 207 &  0 & 0.066 \\
 6 & 18\uph02\upm12\zdot\ups63 & $-30\arcd25\arcm00\zdot\arcs1$ &  407.44 & 133.55 & 18.150 &  5.070 & 13.072 &  11 &  0 & 0.222 & 217 &  0 & 0.102 \\
 7 & 18\uph02\upm11\zdot\ups18 & $-30\arcd24\arcm56\zdot\arcs7$ &  362.03 & 141.64 & 15.598 &  3.058 & 12.535 &   9 &  1 & 0.023 & 223 &  2 & 0.015 \\
 8 & 18\uph02\upm07\zdot\ups45 & $-30\arcd24\arcm54\zdot\arcs9$ &  245.59 & 145.67 & 16.275 &  4.049 & 12.218 &  11 &  0 & 0.059 & 212 &  0 & 0.025 \\
 9 & 18\uph02\upm11\zdot\ups38 & $-30\arcd24\arcm52\zdot\arcs6$ &  368.36 & 151.45 & 15.343 &  3.322 & 12.015 &  10 &  0 & 0.101 & 215 &  0 & 0.054 \\
10 & 18\uph02\upm03\zdot\ups26 & $-30\arcd24\arcm47\zdot\arcs6$ &  114.56 & 163.08 & 14.999 &  3.369 & 11.624 &  11 &  0 & 0.070 & 210 &  0 & 0.031 \\
11 & 18\uph02\upm00\zdot\ups00 & $-30\arcd24\arcm47\zdot\arcs3$ &   12.71 & 163.60 &   --   &   --   & 12.666 &   0 &  0 &   --  & 111 &  3 & 0.012 \\
12 & 18\uph02\upm15\zdot\ups40 & $-30\arcd24\arcm36\zdot\arcs2$ &  493.86 & 191.32 & 18.808 &  5.679 & 13.118 &   6 &  0 & 0.072 & 207 &  0 & 0.073 \\
13 & 18\uph02\upm04\zdot\ups24 & $-30\arcd24\arcm29\zdot\arcs0$ &  144.92 & 207.99 & 14.200 &  2.561 & 11.633 &  10 &  0 & 0.040 & 219 &  0 & 0.027 \\
14 & 18\uph02\upm01\zdot\ups05 & $-30\arcd24\arcm15\zdot\arcs6$ &   45.27 & 240.22 &   --   &   --   & 11.950 &   0 &  0 &   --  & 166 &  0 & 0.009 \\
15 & 18\uph02\upm08\zdot\ups06 & $-30\arcd24\arcm08\zdot\arcs8$ &  264.47 & 257.18 & 15.898 &  2.873 & 13.020 &  10 &  0 & 0.018 & 219 &  0 & 0.014 \\
16 & 18\uph02\upm00\zdot\ups65 & $-30\arcd24\arcm07\zdot\arcs2$ &   32.53 & 260.56 &   --   &   --   & 12.531 &   0 &  0 &   --  & 152 &  0 & 0.012 \\
17 & 18\uph02\upm03\zdot\ups60 & $-30\arcd24\arcm07\zdot\arcs4$ &  124.85 & 260.36 & 16.662 &  3.715 & 12.940 &  10 &  0 & 0.155 & 219 &  0 & 0.015 \\
18 & 18\uph02\upm13\zdot\ups46 & $-30\arcd23\arcm56\zdot\arcs6$ &  433.13 & 287.02 & 16.145 &  3.402 & 12.736 &  11 &  0 & 0.033 & 230 &  0 & 0.020 \\
19 & 18\uph02\upm03\zdot\ups70 & $-30\arcd23\arcm44\zdot\arcs5$ &  127.99 & 315.63 & 14.878 &  2.571 & 12.304 &  11 &  0 & 0.040 & 215 &  0 & 0.015 \\
20 & 18\uph02\upm02\zdot\ups05 & $-30\arcd23\arcm42\zdot\arcs8$ &   76.14 & 319.50 &   --   &   --   & 11.544 &   0 &  0 &   --  & 196 &  0 & 0.014 \\
21 & 18\uph02\upm06\zdot\ups93 & $-30\arcd23\arcm33\zdot\arcs1$ &  228.74 & 343.38 & 15.258 &  2.613 & 12.641 &  11 &  0 & 0.017 & 218 &  0 & 0.013 \\
22 & 18\uph02\upm02\zdot\ups41 & $-30\arcd23\arcm24\zdot\arcs2$ &   87.39 & 364.53 &   --   &   --   & 11.738 &   0 &  0 &   --  & 207 &  0 & 0.054 \\
\noalign{\vskip5pt}
\hline}
\end{landscape}
\newpage

\Section{Data Tests}  
\Subsection{Photometry}  
Quality of the OGLE-II photometry can be assessed from comparison of 
magnitudes of stars located in the overlaps between neighboring fields. 
Because each of the fields was calibrated independently such a comparison  
provides information on accuracy of calibration and the typical accuracy of  
photometry. Figs.~3 and 4 present differences of magnitudes for stars with  
magnitudes brighter than ${I=17}$~mag and ${V=19}$~mag, plotted as a function 
of line number for three fields of different stellar density. The average 
difference of magnitudes is typically below 0.01~mag indicating good 
consistency of the calibration procedure. The typical sigma of the Gaussian 
fitted to the  histogram of differences of magnitudes is about 0.025~mag for 
both, the {\it V} and {\it I}-band.

\Subsection{Completeness}
To estimate the completeness of detection of stars in the OGLE fields we 
performed similar set of tests as in the case of the SMC and LMC maps. For 
details the reader is referred to Udalski \etal (1998). In short, we selected 
a $512\times512$ pixel subframe of the tested field and added artificial stars in 
random locations. Their magnitude distribution is provided in the first column 
of Table~4. The subframe was then reduced by the standard reduction procedure 
and completeness of recovering the artificial stars was studied. One hundred 
such tests were performed for each subframe. 

Table~4 presents results of these tests for three fields of different stellar 
density for {\it I}-band. As can be seen the completeness is high down to 
stars as faint as ${I\approx18.0}$~mag. For fainter stars it gradually drops. 
The completeness is  also a strong function of stellar density of the field. 

\renewcommand{\arraystretch}{1.1}
\MakeTable{c@{\hspace{6pt}}c@{\hspace{6pt}}c@{\hspace{6pt}}
c@{\hspace{6pt}}c@{\hspace{6pt}}c}
{12.5cm}{Completeness of the Galactic bulge maps}
{\hline
Stars &&\multicolumn{3}{c}{Completeness}\\
per bin & $I$  & SC$\_$34 &  SC$\_$40 &  SC$\_$44 \\
\hline
~2 &  12.8 &  99.0 &  98.5 &  99.5 \\ 
~5 &  13.3 &  98.4 &  99.8 &  99.2 \\ 
~7 &  13.8 &  96.1 &  99.1 &  98.7 \\ 
10 &  14.3 &  97.4 &  99.2 &  98.8 \\ 
12 &  14.8 &  96.2 &  98.7 &  98.4 \\ 
15 &  15.3 &  96.8 &  98.6 &  99.1 \\ 
17 &  15.8 &  94.8 &  98.1 &  98.4 \\ 
20 &  16.3 &  93.8 &  96.5 &  97.9 \\ 
22 &  16.8 &  91.4 &  96.0 &  97.3 \\ 
25 &  17.3 &  87.4 &  94.0 &  96.3 \\ 
27 &  17.8 &  79.8 &  91.2 &  93.6 \\ 
30 &  18.3 &  67.3 &  85.2 &  91.0 \\ 
32 &  18.8 &  48.5 &  75.5 &  86.9 \\ 
35 &  19.3 &  26.9 &  56.9 &  80.0 \\ 
\hline}

\Section{Color-Magnitude Diagrams}
Figs.~7--13 show {\it I} \vs $(V-I)$ color-magnitude diagrams (CMDs) of 
selected  Galactic bulge fields. We present these diagrams to illustrate 
quality of data and potential usefulness of the OGLE-II maps for studying 
properties of the Galactic bulge populations. Only part of stars (\ie about 
10--30\% of the total number, depending on the field) from each  field are 
plotted in these figures for clarity. 

The CMDs in Figs.~7--13 clearly show the most characteristic features of 
stellar populations toward the Galactic center regions: the main sequence of 
disk stars (long almost vertical strip at blue side of the CMD), Galactic 
bulge red giant branch (vertical strip at red part of the CMD) with prominent 
red clump, Galactic bulge main sequence turn-off part etc. 

Figs.~7--13 also show how severely the interstellar extinction affects the 
diagrams. In Fig.~7 CMDs of two fields located closest (about $1\zdot\arcd25$) 
to the Galactic center are plotted. In BUL$\_$SC44 field the interstellar 
extinction is so high that only the main sequence of disk stars is visible. In 
BUL$\_$SC5, located somewhat farther from the center, the Galactic bulge red 
giant branch with highly elongated red clump becomes visible, but their large 
parts are still below the detection limit. Fig.~8 presents CMDs of two fields 
located about $1\zdot\arcd8$ from the Galactic center (BUL$\_$SC37 and 
BUL$\_$SC3). The interstellar extinction is much smaller than in the fields in 
Fig.~7 but it still changes significantly across the fields so the Galactic 
bulge red giant branch and red clump are severely smeared. In Fig.~9 CMDs of 
two additional fields from this group (BUL$\_$SC4 and BUL$\_$SC39) are shown. 
As the fields are farther from the Galactic center, the interstellar 
extinction becomes smaller and more uniform, although the variations across 
the fields are still significant. Fig.~10 presents CMDs of two fields located 
about $3\zdot\arcd2$ from the center. The fields (in particular BUL$\_$SC22) 
cover regions with  patches of higher extinction what again smears the red 
giant branch and red clump. CMD of BUL$\_$SC43 field located at positive 
galactic latitude indicates as well high and non-uniform extinction on that 
side of the bulge while that of BUL$\_$SC6 field located farther at negative 
latitudes shows much smaller and very uniform extinction (Fig.~11). Fig.~12 
presents CMDs of two fields located in the Baade's window. Extinction in 
BUL$\_$SC1 is much more uniform than in BUL$\_$SC45. Finally, in Fig.~13  we 
plot CMDs of two fields (BUL$\_$SC2 and BUL$\_$SC26) located much further from 
the Galactic center in the regions of relatively small and uniform 
interstellar extinction. 

\Section{Data Availability}
The {\it VI} maps of the Galactic bulge are available to the astronomical community 
in the electronic form from the OGLE archive: 
\begin{center}
{\it http://www.astrouw.edu.pl/\~{}ogle} \\
{\it ftp://sirius.astrouw.edu.pl/ogle/ogle2/maps/bulge/}\\
\end{center}
or its US mirror
\begin{center}
{\it http://bulge.princeton.edu/\~{}ogle}\\
{\it ftp://bulge.princeton.edu/ogle/ogle2/maps/bulge/}\\
\end{center}

Also {\it I}-band FITS template images of the OGLE-II fields are included.  
Total volume of the compressed data is equal to about 1.8~GB. Usage of the 
data is allowed under the condition of acknowledgment to the OGLE project with 
a reference to this paper. 

We provide these data in the most original form to avoid any additional 
biases. For instance we do not mask bright stars which often produce many 
artifacts, but such a masking could potentially remove some interesting 
information on objects located close to them. We also do not remove objects 
which are located in overlapping areas between the neighboring fields. 
Cross-identification of these objects can be easily done based on provided 
equatorial coordinates. 

\Acknow{We would like to thank Prof.\ Bohdan Paczy\'nski for many discussions  
and help at all stages of the OGLE project. We thank Dr.\ D.S.~Graff for  
information about the systematic error in the early OGLE calibrations. We also  
thank Dr.~A.~Olech for carrying out part of the observations of the Galactic 
bulge in the 1997 observing season. The paper was partly supported by the 
Polish KBN grant 2P03D01418 to M.~Kubiak. Partial support for the OGLE 
project was provided with the NSF grants AST-9820314 and AST-0204908 and NASA 
grant NAG5-12212 to B.~Paczy\'nski. We acknowledge usage of The Digitized Sky 
Survey which was  produced at the Space Telescope Science Institute based on 
photographic data  obtained using The UK Schmidt Telescope, operated by the 
Royal Observatory  Edinburgh.}

\vskip17mm
\centerline{\bf Figure Captions}
\vskip3mm
\begin{figure}[htb]
\FigCap{Differences of magnitudes of the same objects in the overlapping regions 
of fields of low (upper panel), medium (midle panel) and high (bottom
panel) stellar density in the {\it I}-band filter.}
\vskip5mm
\FigCap{Same as Fig.~4 for the {\it V}-band filter.}
\vskip5mm
\FigCap{Standard deviation as a function of magnitude for {\it V} and
{\it I}-band in the highest density field BUL$\_$SC34.}
\vskip5mm
\FigCap{Standard deviation as a function of magnitude for {\it V} and
{\it I}-band in the low density field BUL$\_$SC44.}
\vskip5mm
\FigCap{Color-magnitude diagrams of the fields BUL$\_$SC44 and BUL$\_$SC5.}
\vskip5mm
\FigCap{Color-magnitude diagrams of the fields BUL$\_$SC37 and BUL$\_$SC3.}
\vskip5mm
\FigCap{Color-magnitude diagrams of the fields BUL$\_$SC4 and BUL$\_$SC39.}
\vskip5mm
\FigCap{Color-magnitude diagrams of the fields BUL$\_$SC22 and BUL$\_$SC23.}
\vskip5mm
\FigCap{Color-magnitude diagrams of the fields BUL$\_$SC43 and BUL$\_$SC6.}
\vskip5mm
\FigCap{Color-magnitude diagrams of the fields BUL$\_$SC1 and BUL$\_$SC45.}
\vskip5mm
\FigCap{Color-magnitude diagrams of the fields BUL$\_$SC2 and BUL$\_$SC26.}
\end{figure}
\end{document}